\documentstyle[12pt,epsfig]{article}

\newcommand{\be}{\begin{eqnarray}}
\newcommand{\ee}{\end{eqnarray}}

\title{\begin{flushright}
{\normalsize NUC-MINN-99/4-T\\
March 1999 \\}
\end{flushright}
\vspace*{0.1in}
{\bf SIZE OF FIREBALLS CREATED IN HIGH ENERGY HEAVY ION COLLISIONS
AS INFERRED FROM COULOMB DISTORTIONS OF PION SPECTRA}}
\author{{\bf Alejandro Ayala}$^1$ \\
  {\it Instituto de Ciencias Nucleares UNAM}\\ \vspace*{0.2in}
 {\it A.P. 70-543, M\'exico D.F. 04510, M\'exico}\\
{\bf Sangyong Jeon}$^2$ \\
  {\it Nuclear Science Division}\\
 {\it Lawrence Berkeley National Laboratory} \\ \vspace*{0.2in}
 {\it Berkeley, CA 94720}\\
{\bf Joseph Kapusta}$^3$ \\
  {\it School of Physics and Astronomy}\\
   {\it University of Minnesota}\\ {\it Minneapolis, MN 55455}}

\date{}

\begin{document}

\maketitle
\begin{abstract}

We compute the Coulomb effects produced by an expanding, highly charged
fireball on the momentum distribution of pions.
We compare our results to data on Au+Au at 11.6 A GeV from E866 at
the BNL AGS and to data on Pb+Pb at 158 A GeV from NA44 at
the CERN SPS.  We conclude that the distortion of the spectra at low
transverse momentum and mid-rapidity can be explained in both experiments
by the effect
of the large amount of participating charge in the central rapidity region.
By adjusting the fireball expansion velocity to match the average
transverse momentum of protons, we find a best fit when the fireball
radius is about 10 fm, as determined by the moment when the pions
undergo their last scattering.  This value is common to both the AGS and
CERN experiments.
\end{abstract}

\begin{flushleft}
\end{flushleft}

%\vspace*{0.1in}

\begin{flushleft}
$^1$ ayala@xochitl.nuclecu.unam.mx\\
$^2$ jeon@nta2.lbl.gov\\
$^3$ kapusta@physics.spa.umn.edu\\
\end{flushleft}

\newpage

The purpose of colliding heavy nuclei at relativistic and ultrarelativistic
energies is to produce high density, high temperature matter, either
hadronic or quark-gluon plasma. One way of inferring the properties of matter
under these extreme conditions is to compare the single particle momentum
distributions of secondaries to those obtained in collisions of lighter
systems at similar beam energies. An example is provided by the qualitative
differences found between the transverse momentum distributions
from heavy-ion reactions and those from p--p collisions. In the former, the
bulk features of the spectrum can successfully be explained by
the large amount of secondary scattering, potentially leading to the
onset of hydrodynamical behavior \cite{qm96}.

Another feature to account for in the collision of heavy
systems is the presence of a large amount of electric charge that will
influence the dynamics of secondary charged particles. Due to
the long-range nature of the Coulomb field, the spectrum of charged
particles can still be distorted even after freeze-out. For central
collisions this Coulomb effect can be more significant in the case of strong
stopping when the participant charge in the central rapidity region is an
important fraction of the initial charge, particularly on the low
momentum particles.

In a recent paper \cite{AK} two of us explored the
influence of the time dependent electric field produced by an expanding,
highly charged fireball on the low energy part of the spectrum of charged
test particles, taken to be kaons \cite{otherkaon}.
The analysis was based on the solution to
Vlasov's equation assuming that the test particles' dynamics was
nonrelativistic. In this paper we extend that analysis, treating
the test particle relativistically. We also apply the analysis to the
transverse momentum distribution of charged pions and compare the
results to recent measurements of Au+Au collisions by the E866
collaboration at the BNL AGS and to recent measurements of
Pb+Pb collisions by the NA44 collaboration at the CERN SPS.
Relativistic calculations for pions and kaons have also been done
in somewhat different models by Barz {\it et al.} \cite{Barz}.

We briefly recall the basics of the model used in Ref.~\cite{AK} and how it
is modified for relativity.  A uniformly charged sphere which has a
total charge $Ze$ and whose radius $R$
increases linearly with time $t$ from a value $R_0$ at time $t_0$ at a
constant surface speed $v_s$ produces an electric potential
\be
  V(r,t)=\left\{ \begin{array}{ll}
  Ze/4\pi r \,\,,   &   r \geq R = v_st \\
  Ze(3R^2-r^2)/8\pi R^3 \,\,,   &   r \leq R = v_st
  \end{array}
  \right. \, .
  \label{eq:potential}
\ee
In the center-of-mass frame of the fireball the charge moves radially
outwards, hence there is no preferred direction and consequently
the magnetic field produced by this moving charge configuration vanishes.
The fireball parameters are related by $R_0=v_st_0$.
If $f^{\pm}({\bf r},{\bf p},t)$ represents the
$\pm e$ test particle phase space distribution then, when ignoring
particle collisions after decoupling, its dynamics is governed by
Vlasov's equation.
\be
   \left[ \frac{\partial}{\partial t} +
   \frac{{\bf p}}{E_p}\cdot\nabla_r \pm
   e{\bf E}({\bf r},t)\cdot\nabla_p \right]
   f^{\pm}({\bf r},{\bf p},t)=0\, ,
   \label{eq:Vlasov}
\ee
where $E_p=\sqrt{p^2+m^2}$, $m$ is the meson's mass and
\be
  e{\bf E}({\bf r},t) = -e\nabla_r V(r,t)=
   \left\{ \begin{array}{ll}
   \left( t_s/4t^3 \right){\bf r}\,\,, & r \leq R = v_st \\
   \left( t_sv_s^3/4r^3\right){\bf r}\,\,, & r \geq R = v_st
   \end{array}
   \right.
   \label{eq:Efield}
\ee
is the time-dependent electric field corresponding to the potential
$V(r,t)$.  Here we have defined the characteristic time $t_s$ by
\be
 t_s\equiv \frac{Ze^2}{\pi mv_s^3} \,  ,
   \label{chartime}
\ee
and we work in units where $c = 1$.  The set of
Eqs.~(\ref{eq:potential})--(\ref{chartime}) are
relativistically correct.  Notice that there is no problem with retardation
effects since what is specified in these equations is the electric field and
{\it not} the charge density. Of course, the relationship between the charge
density and the eletric field gets modified due to the finite speed of light,
but we are not interested in that relationship here. The only difference
between the equations to be solved here and those solved in Ref.~\cite{AK}
involve the energy $E_p$ in Eq.~(\ref{eq:Vlasov}) which was approximated by
$m+p^2/2m$ in the nonrelativistic case.

The solution to Eq.~(\ref{eq:Vlasov}) is found by the method of
characteristics. This involves solving the classical equations of motion
and using the solutions to evolve the initial distribution, assumed
here to be thermal
\be
f^{\pm}({\bf r},{\bf p},t_0)
=
\exp\left\{ -\left( E_p \pm V(r,t_0) \right)/T \right\}
\ee
forward in time.
The equations of motion are:
\be
   \frac{d{\bf r}}{dt}&=&\frac{{\bf p}}{E_p}\nonumber\\
   \frac{d{\bf p}}{dt}&=&\pm e{\bf E}({\bf r},t)\, .
   \label{eq:eqsofmotion}
\ee
The pion's asymptotic momentum is calculated numerically by a 6'th order
Runge-Kutta method with adaptive step sizes from a set of initial
phase-space positions. The final momentum distribution is the result
of computing the trajectories for many initial phase
space points.  The initial radial position was incremented in
N$_r$=50 steps with spacing $\Delta r/R_0$ = 0.02. The initial
momentum was incremented in N$_p$=600 steps with spacing
$\Delta p$ = 1 MeV.  The cosine of the angle between the initial
position and momentum vectors was incremented in N$_z$=100 steps
of size $\Delta \cos\theta=0.02$.  Hence the total
number of trajectories computed was $3\times 10^6$
for each set of initial conditions.

For central Pb+Pb collisions at 158 A GeV at the SPS, for which NA44
and NA49 have reported a significant amount of
stopping \cite{NA44,NA49}, the number of protons per unit rapidity in
the central region is on the order of 30. If we consider that the
rapidity region spanned by the fireball is between 1 and 5 units,
then we take as the effective fireball's charge $Z=120$.

To estimate the surface velocity $v_s$ we note that the {\it average}
transverse velocities of pions and protons in Pb+Pb collisions
at the SPS are greater than 0.6 \cite{qm96}.  Therefore, to represent the
electric field it is reasonable to take the {\it maximum} velocity
of charges to be greater than that; we shall use $v_s = 0.8$ and
$v_s = 1$.

\vspace{-0.5in}
\begin{figure}[h!]
\centerline{\epsfig{file=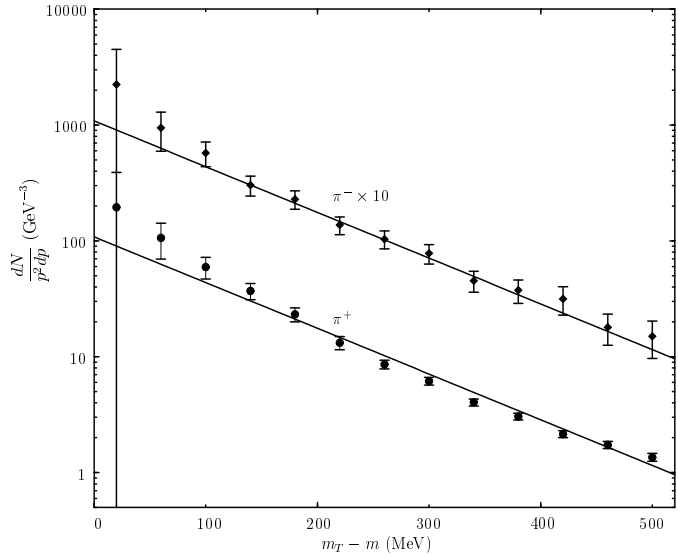,height=6in,width=4.5in}}
\vspace*{-2in}
\caption{ The transverse mass distributions at mid-rapidity
          for $\pi^+$ and $\pi^-$.  The solid lines are exponentials
          with an inverse slope of 110 MeV.  The data are from 
          NA44~\cite{NA44qm}. }
\end{figure}

The model we are using is spherically symmetric.  It has long been
known \cite{KLM} that the momentum distributions at the SPS are somewhat
forward-backward peaked, even for central Pb+Pb collisions \cite{Jones}.
A spherically symmetric model is, of course, not essential to the basic physics
and can be relaxed at the expense of additional computing time.
Nevertheless, we can provide the following arguments in favor of its use.  
First,
we will be comparing with the transverse momentum distributions at mid-rapidity
where the impact of spherical asymmetry should be less important than near the
fragmentation regions.  Second, pion interferometry of central Au+Au collisions
at the AGS \cite{Barrette} and of central Pb+Pb collisions at the SPS
\cite{Wiedemann} both yield comparable values for the transverse and
longitudinal radii at the time of pion freezeout or strong decoupling.
These are about twice the radius of a cold gold or lead nucleus.  Third,
we shall show later that the transverse surface of the fireball must expand
outwards with a speed more than 90\% that of light in order to reproduce the
average proton transverse momentum.  Since the longitudinal surface of the
fireball cannot travel faster than the speed of light, this means that in
velocity space the fireball is nearly symmetric.  These phenomena, although not
yet measured at that time, were already known to Landau~\cite{Landau}. His
model envisioned the two nuclei stopping each other within one Lorentz
contracted nuclear diameter, then undergoing a one dimensional expansion along
the beam axis.  After a time of order (nuclear radius)/c
the longitudinal size of the fireball had become
comparable to the transverse size and, furthermore, the rarefaction wave would
have caused transverse expansion to set in.  As a rough approximation one
therefore could imagine the original one dimensional expansion to become three
dimensional after that time.  This model was later developed by others,
including Cooper, Frye and Schonberg \cite{Cooper}.  The essential insight from
Landau's model is not the degree of stopping but rather the point that
significant transverse expansion sets in after the longitudinal and transverse
radii become comparable in magnitude.  Thereafter, from the point of
view of a distant observer, the
expansion is not as asymmetric as one might originally think.

\vspace{-0.5in}
\begin{figure}[h!]
\centerline{\epsfig{file=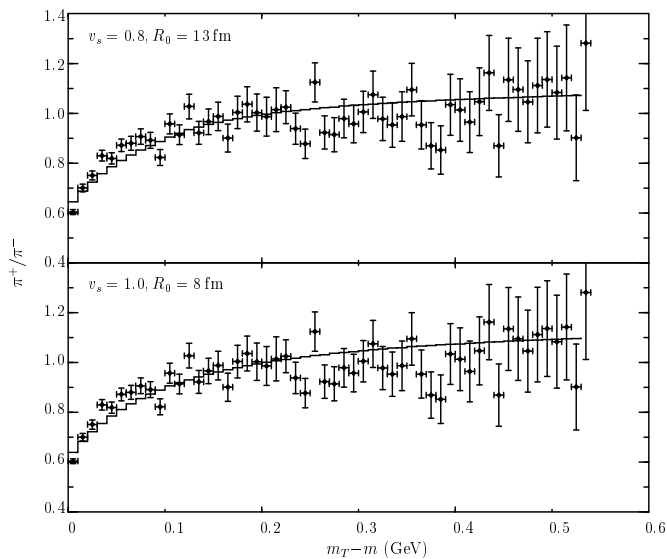,height=6in,width=4.5in}}
\vspace*{-2in}
\caption{ The ratio $\pi^+/\pi^-$ vs. $m_T - m$ at the SPS.  The data is
          from \cite{NA44}. Top panel: The curve corresponds to the Coulomb
          calculation described in the text with surface velocity 0.8 and
          initial radius 13 fm. Bottom panel: Same as top panel but with the
          parameters 1 and 8 fm, respectively. }
\end{figure}

The collaboration NA44 has reported a suppression of the $\pi^+/\pi^-$ ratio
as $p_T \rightarrow 0$ in Pb+Pb collisions, presumably due to the
Coulomb field of the fireball.  Only a very small, or no, suppression
is observed in S+Pb collisions.  To model the primordial distribution
we use an exponential parametrization of $dN/p^2dp$. Fig.~1 shows
this representation \cite{ECHX} in comparison to the
transverse mass distributions of positive and negative pions \cite{NA44qm}
with a value of $T_{\rm eff} = 110$ MeV.  This common
fit (apart from the absolute normalization) is then used to model
the {\it initial} pion distribution, and the Coulomb computation
we do is a perturbative correction to it.  In essence we sprinkle
the pions around at time $t_0$ according to a relativistic Boltzmann
distribution with an effective temperature of 110 MeV and solve
the Vlasov equation for subsequent time. This assumes that all
resonances that can decay into charged pions have already done so,
hence their effects are contained in this parametrization.
Even most of the relatively long-lived omega-mesons should have decayed
by the this late freeezeout time and relatively low temperature.

The procedure we follow is to fix $v_s$ at either 0.8 or 1 and then
vary $R_0$ to get the best fit to the ratio $\pi^-/\pi^+$ as a function
of $m_T-m$ in the range from 0 to 500 MeV.  The data's
bin size is 10 MeV.  The data are shown in Fig. 2 and compared to
the results of calculations with $v_s=0.8$, $R_0=13$ fm and
$v_s=1$, $R_0=8$ fm.  There is no visual difference between the
calculations with the two parameter sets.  The reason the
final Coulomb corrected ratio is nearly identical for these two
sets of parameters is easily understood.  According to the study
in Ref.~\cite{AK}, the longer the test charge stays in the electric field
of the fireball the more it is modified from its initial form.
Thus, for the more slowly expanding fireball ($v_s=0.8$) the radius
must be larger (13 vs. 8 fm) in order to reduce the magnitude of
the electric field so as to represent the data.

\vspace{-0.5in}
\begin{figure}[h!]
\centerline{\epsfig{file=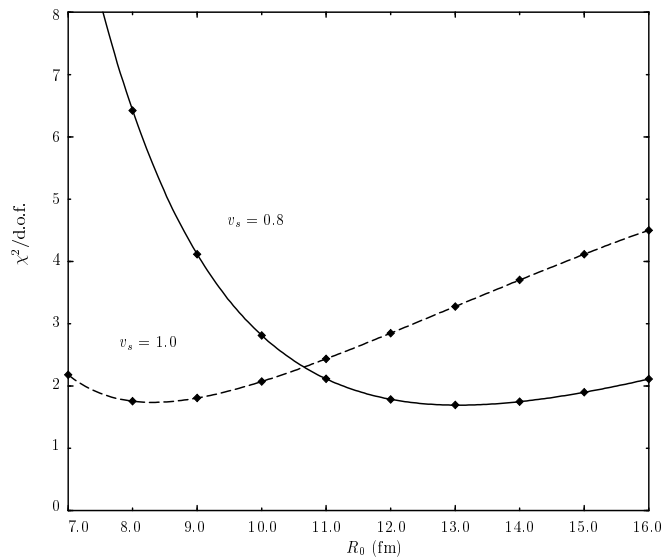,height=6in,width=4.5in}}
\vspace*{-2in}
\caption{ Chi-squared vs. radius for fixed values of the surface
          velocity for pions for central Pb+Pb collisions at the SPS.  
          The curves represent polynomials which pass through 
          the values of $R_0$ for which computations were done. }
\end{figure}

In Fig. 3 we plot the chi-squared for the $\pi^+/\pi^-$ ratio as a
function of radius $R_0$ for the surface velocity $v_s$ fixed
at 0.8 and 1.  The minimum is achieved at about 13 and 8.3 fm,
respectively.  The difference in the chi-squared for the two
values of $v_s$ is not significant enough to make a judgement
on which gives the better fireball physics.  The reason for the
close coupling of the two parameters, $v_s$ and $R_0$, is explained
by recognizing that it is the ratio of the characteristic expansion
time $t_0$ to the characteristic Coulomb time $t_s$ which is the
relevant quantity.  Since $t_0/t_s \propto v_s^2 R_0$ we would
expect comparable fits to the data will be obtained when
$R_0$ varies inversely with the square of $v_s$; this is borne
out by our explicit numerical calculations.
We remark that a different parametrization of the
primordial pion distribution, such as a two temperature fit, might
well lead to an even better representation of the data.

\vspace{-0.8in}
\begin{figure}[h!]
\centerline{\epsfig{file=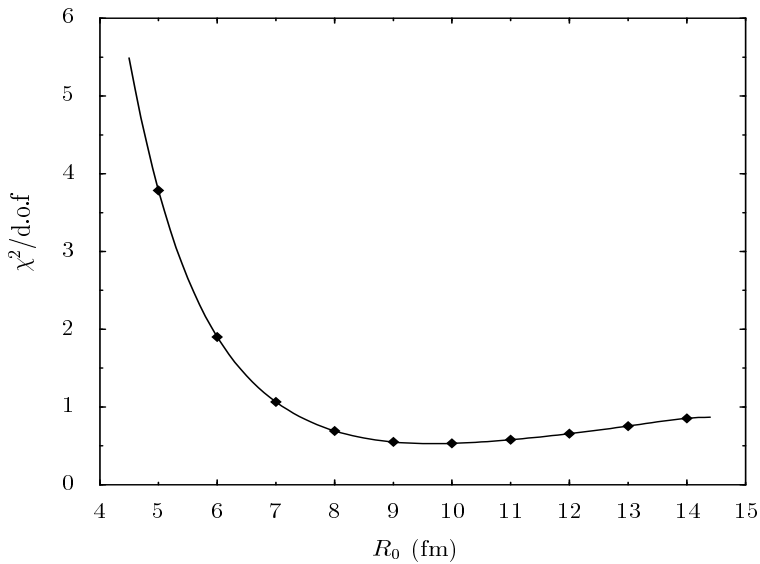,height=8in,width=7in}}
\vspace*{-4in}
\caption{ Chi-squared vs. radius for a surface velocity of 0.914 for pions for
          central Au+Au collisions at the AGS. The curves represent polynomials
          which pass through the values of $R_0$ for which computations were
          done. }
\end{figure}

The immediate question now is how to determine the appropriate
value of $v_s$.  We shall do this by matching the average transverse
momentum of the protons as inferred from our assumption of a uniformly
expanding sphere with the measured one.  The model gives
\begin{equation}
\langle p_T \rangle = \frac{\int d^3r \, \rho(r,t) \, \sin\theta
\, m_P \, v/\sqrt{1-v^2}}{\int d^3r \, \rho(r,t)} \, ,
\end{equation}
where $\rho(r,t) = \rho_0 (t_0/t)^3 \Theta(v_st-r)$ and
$v=r/t$.  The result is:
\begin{equation}
\langle p_T \rangle = \pi \, \frac{2-(2+v_s^2)
\sqrt{1-v_s^2}}{4v_s^3} \, m_P \, .
\end{equation}
In the limits $v_s \rightarrow 0$ and $v_s = 1$ one gets
$\langle p_T \rangle = 3\pi m_P v_s/16$ and
$\langle p_T \rangle = \pi m_P/2$, respectively.  Note that $v_s$
should {\it not} be interpreted as a hydrodynamic flow velocity.
Rather, it embodies the combined effects of hydrodynamic flow and
thermal motion of the net charge carriers, mainly protons.  Both
NA44 \cite{NA44qm} and NA49 \cite{NA49qm} have
reported the transverse mass distribution in central
Pb+Pb collisions at midrapidity to be $dN/p_Tdp_T \propto
\exp(-m_T/T_P)$ with $T_P$ = 290 MeV.  This corresponds
to an average transverse momentum of 825 MeV.
The value $v_s$ = 0.916 gives the best matching of
the model to the proton spectra.  This finally pins down
the freezeout radius for pions to be 10 fm.

\vspace{-0.8in}
\begin{figure}[h!]
\centerline{\epsfig{file=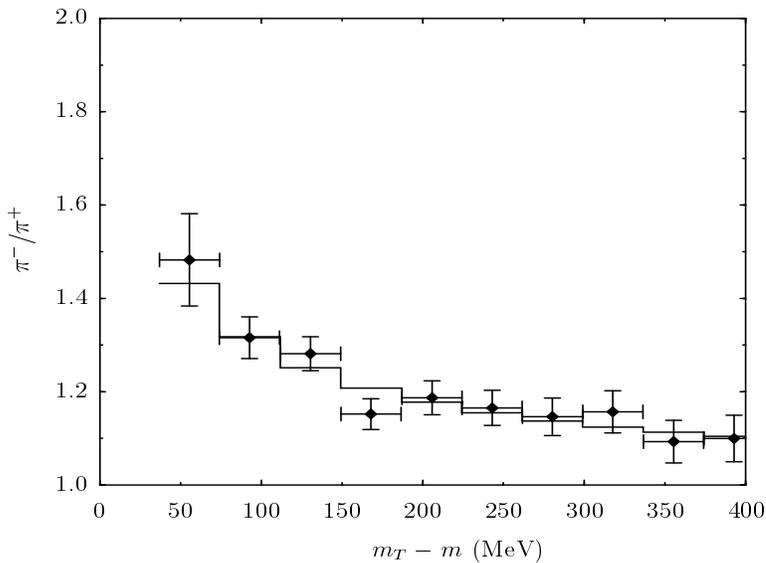,height=8in,width=7in}}
\vspace*{-4in}
\caption{ The ratio $\pi^-/\pi^+$ vs. $m_T - m$ at the AGS.  The data is
          from \cite{Ahle}.  The fireball radius at pion freezeout is 10 fm. }
\end{figure}

We now repeat the above analysis for the BNL experiment E866 \cite{Ahle}.
This experiment measured protons and positive and negative pions for Au+Au
collisions at 11.6 A GeV.  We focus on the most central events, those with the
highest multiplicity and transverse energy.  The average proton transverse
momentum is 820 MeV yielding the surface speed $v_s = 0.914$.  The number of
participating protons maybe be estimated from the data to be $Z = 116$.  The
slopes of the $\pi^+$ and $\pi^-$ are the same to within several MeV; their
average is 111 MeV.  All of these quantities are remarkably similar to those in
central Pb collisions at the much higher energy of the SPS.

The experiment reported the ratio $\pi^-/\pi^+$ for a central window of
rapidity.  We calculated the ratio with two parameters: the radius at pion
freezeout $R_0$ and the absolute normalization of the individual spectra.  The
chi-square per degree of freedom is shown in Fig. 4 as a function of $R_0$.  It
has a minimum of 0.53 near 10 fm, the same as Pb collisions at the much higher
SPS energy!  The lower value of chi-square in this case as compared to the SPS
data may be due to the better approximation of spherical symmetry.  The
calculated ratio is compared to the data in Fig. 5 and its appearance is quite
satisfactory.

In conclusion, we have demonstrated that the suppression of the
ratio $\pi^+/\pi^-$ in central Pb+Pb collisions at the SPS
as observed by NA44 and in central
Au+Au collisions at the AGS as observed by E866 can be quantitatively
understood as a Coulomb effect.
The large electric fields generated by the expanding fireballs seem to
distort the pion ratios at small transverse momenta and midrapidity
in accordance with these experiments.
In principle these ratios should provide a very good
measure of the size of the fireball at the time pions cease
collisions and begin to free-stream.  From
experiments \cite{qm96} one may estimate that about 2500 baryons
and mesons are produced with rapidities within 2 units of mid-rapidity in the
central Pb+Pb collisions.
Associating these with a fireball of radius 10 fm means an average
density in the fireball reference frame of 0.60/fm$^3$; this is
4 times nuclear density!  Taking into account relativistic length
contraction the local density is smaller by a factor
\begin{equation}
\frac{3}{R_0^3} \int_0^{R_0} dr\,r^2\,\gamma(r) =
\frac{3}{2v_s^2}\left[ \sin^{-1}(v_s) - v_s \sqrt{1-v_s^2}\right] \, .
\end{equation}
For $v_s = 0.916$ this implies a local density of
0.42/fm$^3$, still quite considerable.  This implies
that at the moment that free-streaming of mesons begins the hadronic
density is still interestingly high.

To make further progress it is clear that more accurate data are
called for but, more importantly, one should also compute the influence
of the Coulomb force within an event simulator which takes into account a more
accurate description of the space-time evolution of the collision.

\section*{Acknowledgements}
The authors wish to thank Nu Xu, P. Jacobs and J. Pisut for insightful
comments and suggestions.
This work was supported by the U.S. National Science Foundation under
grant NSF PHY94-21309, by the U.S. Department of Energy under grants
DE-FG02-87ER40328, DE-AC03-76SF00098 and DE-FG03-93ER40792,
and by the CONACyT-M\'exico under grant I27212-E.


\begin{thebibliography} {20}

\bibitem{qm96} Good overviews are provided by the proceedings of the
Quark Matter conferences, the most recent in print being Quark Matter
'97, Nucl. Phys. {\bf A638}, (1998).

\bibitem{AK}
A. Ayala and J. Kapusta, Phys. Rev. C {\bf 56}, 407 (1997).

\bibitem{otherkaon}
Similar studies were carried out independently by V. Koch, Nucl. Phys.
{\bf A590}, 531c (1995).

\bibitem{Barz} H. W. Barz, J. P. Bondorf,
J. J. Gaardh{\o}je, and H. Heiselberg, Phys. Rev. C {\bf 56}, 1553 (1997);
{\it ibid}. {\bf 57}, 2536 (1998).

\bibitem{NA44} NA44 Collaboration, H. B{\o}ggild {\it et al}.,
Phys. Lett. {\bf B372}, 339 (1996).

\bibitem{NA49}
NA49 Collaboration, T. Wienold {\it et al}.,
Nucl. Phys. {\bf A610}, 76c (1996).

\bibitem{KLM} KLM Collaboration, H. von Gersdorff {\it et al}.,
Phys. Rev. C {\bf 39}, 1385 (1989).

\bibitem{Jones} NA49 Collaboration, P. G. Jones {\it et al}.,
Nucl. Phys. {\bf A610}, 188c (1996).

\bibitem{Barrette} E877 Collaboration, J. Barrette {\it et al}., Nucl. Phys.
{\bf A610}, 227c (1996).

\bibitem{Wiedemann} U. Wiedemann, B. Tomasik and U. Heinz, Nucl. Phys. {\bf
A638}, 475c (1998).

\bibitem{Landau} L. D. Landau, Izv. Akad. Nauk SSSR (Physics Series) {\bf 17},
51 (1953); S. Z. Belenkij and L. D. Landau, Nuovo Cim. Suppl. {\bf 3}, 15
(1956).

\bibitem{Cooper} F. Cooper and G. Frye, Phys. Rev. D {\bf 10}, 186 (1974); F.
Cooper, G. Frye and E. Schonberg, {\it ibid.} {\bf 11}, 192 (1975).

\bibitem{ECHX}
It is also possible to represent transverse momentum distributions
as arising from a combination of temperature and hydrodynamic flow,
as illustrated more recently by S. Esumi, S. Chapman, H. van Hecke
and N. Xu, Phys. Rev. C {\bf 55}, R2163 (1997).  What matters most in
the Coulomb analysis is the momentum distribution of mesons at the
moment of last collision.  We have investigated what happens when
a combination of flow velocity and temperature is used to represent
the kaons' momentum distributions and it makes very little difference.
In fact, the only difference arises from the spatial distribution of
mesons at the moment of last collision.  Such fine detail is outside
the scope of this study.

\bibitem{NA44qm}
NA44 Collaboration, N. Xu {\it et al}., Nucl. Phys. {\bf A610}, 175c (1996).
There is a slight difference in the numerical value
of $T_{\rm eff}$ depending on whether it is $d^3N/d^3p$ or
$E d^3N/d^3p$ which is fit with an exponential.

\bibitem{NA49qm} NA49 Collaboration, G. Roland {\it et al}., Nucl Phys. {\bf
A638}, 91c (1998).

\bibitem{Ahle} E866 Collaboration, L. Ahle {\it et al}., Phys. Rev. C {\bf
57}, R466 (1998).
\end{thebibliography}
\end{document}